# Solution of the wave equation in a tridiagonal representation space

E. El Aaoud[1], H. Bahlouli[2], and A. D. Alhaidari[3]

[1]*Physics Department, University of Hail, P.O. Box 2440 Hail, Saudi Arabia*
[2]*Physics Department, King Fahd University of Petroleum & Minerals, Dhahran 31261, Saudi Arabia*
[3]*Shura Council, Riyadh 11212, Saudi Arabia*

We use variable transformation from the real line to finite or semi-infinite spaces where we expand the regular solution of the 1D time-independent Schrödinger equation in terms of square integrable bases. We also require that the basis support an infinite tridiagonal matrix representation of the wave operator. By this requirement, we deduce a class of solvable potentials along with their corresponding bound states and stationary wavefunctions expressed as infinite series in terms of these bases. This approach allows for simultaneous treatment of the discrete (bound states) as well as the continuous (scattering states) spectrum on the same footing. The problem translates into finding solutions of the resulting three-term recursion relation for the expansion coefficients of the wavefunction. These are written in terms of orthogonal polynomials, some of which are modified versions of known polynomials. The examples given, which are not exhaustive, illustrate the power of this approach in dealing with 1D quantum problems.

## 1. Introduction

Since the advent of quantum mechanics, exactly solvable potentials were of great interest, both for their mathematical intrigue and for testing the validity of perturbative, numerical and semi-classical approximations of physical systems. Exact solvability of a given Hamiltonian with its boundary conditions entails the exact knowledge of all its eigenfunctions and the corresponding energy spectrum [1]. However, since the early days of quantum mechanics the number of exactly solvable problems is very limited. But, they played a major role in putting the theory on firm grounds and in understanding many physical phenomena [2]. In recent year there have been some efforts in classifying all types of solvable problems based on symmetry considerations. First, the idea of shape invariance played a major role in classifying exactly solvable nonrelativistic quantum problems in distinct classes. Second, new methods were developed to generate solutions of solvable models such as supersymmetric quantum mechanics, potential algebra, path integration, and point canonical transformations [3]. Solvable potentials can be classified into exactly solvable, conditionally-exactly solvable [3,4] or quasi-exactly solvable [5,6]. Exactly solvable potentials are those for which one can determine the whole spectrum analytically for a continuous range of values of the potential parameters. Conditionally-exactly solvable potentials are those for which a solution can be generated only for specific values of the potential parameters [3,4] whereas, quasi-exactly solvable [5,6] potentials are those for which one can determine only part of the spectrum. These developments were carried out over the years by many authors where several classes of these solutions were accounted for and tabulated (see, for example, the references cited in [1]). In those developments, the main objective was to find solutions of the eigenvalue wave equation $H|\Psi\rangle = E|\Psi\rangle$, where $H$ is the Hamiltonian and $E$ is the energy which is either discrete (for bound states) or continuous (for scattering states). Hence one ends up looking for a representation in which the Hamiltonian has a diagonal structure exhibiting the eigenvalues or the spectrum of the associated potential.

In this article, we search for the regular algebraic solution of the one-dimensional wave equation. Hence, we expand the wave function $\Psi$ in the space of square integrable discrete basis elements $\{\phi_n\}_{n=0}^{\infty}$. That is, the wave function is expandable as $|\Psi(x,E)\rangle = \sum_n f_n(E)|\phi_n(x)\rangle$, where $x$ is the whole real line. The basis functions must be compatible with the domain of the Hamiltonian and satisfy the boundary conditions of the problem. However, the main contribution of this work is in relaxing the usual restriction of a diagonal matrix representation of the eigenvalue wave equation. We only require that it be tridiagonal and symmetric. That is, the action of the wave operator on the elements of the basis is allowed to take the general form $(H-E)|\phi_n\rangle \sim |\phi_n\rangle + |\phi_{n-1}\rangle + |\phi_{n+1}\rangle$. For example, in an orthogonal basis, the matrix elements of the wave operator becomes

$$\langle\phi_n|H-E|\phi_m\rangle = (a_n - E)\delta_{n,m} + b_n\delta_{n,m-1} + b_{n-1}\delta_{n,m+1}, \qquad (1.1)$$

where the coefficients $\{a_n, b_n\}_{n=0}^{\infty}$ are real and, in general, functions of the potential parameters. The hope is that by relaxing the diagonal constraint, the space of representations becomes large enough to incorporate a larger class of solvable potentials. Moreover, this approach embodies powerful tools in the analysis of solutions of the wave equation by exploiting the intimate connection and interplay between tridiagonal matrices and the theory of orthogonal polynomials. In such analysis, one is at liberty to employ a wide range of well established methods and numerical techniques associated with these settings such as quadrature approximation and continued fractions [7]. Additionally, since tridiagonal matrices have special and favorable treatments in numerical routines (e.g., in computing their eigenvalues and eigenvectors), the accuracy and convergence of numerical computations are also enhanced.

The matrix wave equation, which is obtained by expanding $|\Psi\rangle$ as $\sum_m f_m |\phi_m\rangle$ in $(H-E)|\Psi\rangle = 0$ and multiplying on the left by $\langle\phi_n|$, results in the following three-term recursion relation (for an orthogonal basis)

$$E f_n = a_n f_n + b_{n-1} f_{n-1} + b_n f_{n+1}. \qquad (1.2)$$

Consequently, the problem translates into finding solutions of the recursion relation for the expansion coefficients of the wave function $\Psi$. In most cases this recursion is solved easily and directly by correspondence with those of well known orthogonal polynomials. It is obvious that the solution of (1.2) is obtained modulo an overall factor which is a function of the physical parameters of the problem but, otherwise, independent of $n$. If we decide to choose the typical normalization where we take the initial seed $f_0 = 1$, then Eq. (1.2) gives $f_n(E)$ as a polynomial of degree $n$ in $E$. Therefore, a solution of the problem is obtained whence the expansion coefficients $\{f_n\}_{n=0}^{\infty}$ are determined. To be mathematically rigorous, one has to prove that the infinite sum $\sum_{n=0}^{\infty} f_n |\phi_n\rangle$ converges. However, the present setting is not proper to carry out such analysis. It might just be sufficient to give a hand-waving argument by pointing out that the completeness of the basis $\{\phi_n\}_{n=0}^{\infty}$ as well as the set of orthogonal polynomials $\{f_n\}_{n=0}^{\infty}$ guarantee such requirement. It should also be noted that the solution of the problem as depicted by Eq.

(1.2) above is obtained for all *E*, the discrete as well as the continuous. Moreover, the representation equation (1.1) clearly shows that the discrete spectrum is easily obtained by diagonalization which requires that:

$$b_n = 0, \text{ and } E = a_n.$$ (1.3)

for all *n*.

In section 2 we present the theoretical formulation of the problem and explain the general approach we adopted in finding solvable potentials. In sections 3-5 we consider few explicit coordinate transformations and derive the corresponding solvable potentials. Interestingly, we rediscover most of the well known solvable potentials such as the generalized Morse, Rosen-Morse, Pöschl-Teller and harmonic oscillator-inverse square potentials. These investigations do not exhaust the set of all 1D problems that are solvable using this approach. In section 6 we present our conclusion and discuss possible extensions of this work and its potential application in treating the J-matrix of scattering in one dimension. Useful formulae and relations satisfied by orthogonal polynomials that are relevant to our present work are given in the appendix.

## 2. Theoretical Formulation

In this section, we present our general strategy in deriving exactly solvable potentials. We consider the 1D time-independent Schrödinger equation

$$\left[ -\frac{\hbar^2}{2m} \frac{d^2}{dx^2} + V(x) \right] \Psi(x,E) = E \, \Psi(x,E) \quad ; \quad x \in \mathbb{R}$$ (2.1)

We would like to transform the above equation to either a finite or semi-infinite space with new coordinate *y*(*x*) while maintaining the boundary conditions. Without loss of generality, we limit our study to the semi-infinite $y \in [0,\infty[$ or finite $y \in [-1,+1]$ configuration spaces. In the following, we will use some specific transformations to bring the variable *x* to either one of these two ranges. Thereafter, by using a discrete $L^2$ basis set we write down the matrix representation of the wave operator $H - E$ and require it to be tridiagonal. Subsequently, we deduce the possible types of potentials that allow for exact solutions within our scheme. The basis function set that will be used in the new variable *y* are either of the Laguerre or Jacobi type corresponding to the semi-infinite or finite spaces, respectively. The Laguerre basis is defined as:

$$\phi_n(y) = A_n y^\alpha e^{-y/2} L_n^\nu(y), \text{ with } A_n = \sqrt{\lambda \, \Gamma(n+1) / \Gamma(n+\nu+1)}$$ (2.2)

where $\alpha$ and $\nu$ are real parameters with $\nu > -1$ and $\alpha \geq 0$ to ensure convergence of the Laguerre polynomial and compatibility with the boundary conditions when the new variable *y* span the semi-infinite interval $[0,\infty[$. The second is called the Jacobi basis and has the following structure:

$$\phi_n(y) = A_n (1+y)^\alpha (1-y)^\beta P_n^{(\mu,\nu)}(y), \, A_n = \sqrt{\frac{\lambda(2n+\mu+\nu+1)\Gamma(n+1)\Gamma(n+\mu+\nu+1)}{2^{\mu+\nu+1}\Gamma(n+\nu+1)\Gamma(n+\mu+1)}}$$ (2.3)

where $\alpha, \beta \geq 0$, $\mu, \nu > -1$. This basis is used when the new variable *y* spans a finite range, which without loss of generality is be taken to be [−1,+1]. Under a suitable coordinate transformation *y*(*λx*), where *λ* is a positive parameter having dimensions of inverse length, the original Schrödinger equation (2.1) becomes

$$\left[ -(y')^2 \frac{d^2}{dy^2} - y'' \frac{d}{dy} + U(y) - \varepsilon \right] \Psi(y,\varepsilon) = 0, \tag{2.4}$$

$$U(y) = V(x(y))/E_0 , \quad \varepsilon = E/E_0 , \quad E_0 = \hbar^2 \lambda^2 / 2m$$

where the energy is measured in units of $E_0$ and the prime on $y$ stands for the derivative with respect to $\lambda x$. Using the fact that the wavefunction can be expanded in $L^2$ basis set as $\Psi(y,E) = \sum_{n=0}^{\infty} f_n(E) \phi_n(y)$, the wave equation reduces to

$$\sum_{n=0}^{\infty} f_n(\varepsilon) \left[ -(y')^2 \frac{d^2}{dy^2} - y'' \frac{d}{dy} + U(y) - \varepsilon \right] \phi_n(y) = 0. \tag{2.5}$$

The use of the differential, recursion, and orthogonality properties of the Laguerre and Jacobi polynomials (shown in the Appendix) in this equation will be very valuable throughout this work for obtaining the fundamental three-term recursion relation for the expansion coefficients of the wavefunction.

It is inherent in our approach that the stationary wavefunction can be expanded in the chosen $L^2$ basis; in the ket notation, $|\Psi\rangle = \sum_{n=0}^{\infty} f_n |\phi_n\rangle$. Then the original Schrödinger equation is required to take a tridiagonal form

$$J_{n,n} f_n + J_{n,n+1} f_{n+1} + J_{n,n-1} f_{n-1} = 0 \tag{2.6}$$

where $J_{m,n} = \langle \phi_m |(H-E)| \phi_n \rangle$ are elements of the "J-matrix" which plays an important role in the J-matrix method of scattering [8]. This tridiagonal requirement is the basis of our approach and establishes the desired connection with orthogonal polynomials whose orthogonality and spectral properties will be of great use in our method.

In the following three sections, we apply the general approach mentioned above to some specific types of variable transformation that leads to analytic solutions of the 1D Schrödinger equation.

## 3. The Oscillator Transformation

It is defined through the transformation $y = (\lambda x)^2$. In this case the $y$ space is $[0,\infty[$, which is a folding of the $x$ configuration space, and the original Schrödinger equation becomes

$$\left[ -4y \frac{d^2}{dy^2} - 2 \frac{d}{dy} + U(y) - \varepsilon \right] \Psi(y,\varepsilon) = 0 \quad ; \quad y \in [0,\infty[ \tag{3.1}$$

The integration measure is defined by[+] : $\int_{-\infty}^{+\infty} dx \cdots = \int_0^{+\infty} \frac{dy}{\lambda \sqrt{y}} \cdots$. Using the differential equation and differential formula of the Laguerre polynomials in the basis (2.2), we

---

[+] We assume that the integrand is a function of $x^2$ (i.e., $y$) which is true for our basis and the sought-after potentials. However, for integrands that are functions of $x$, then we must have:
$$\int_{-\infty}^{+\infty} f(x) dx = \frac{1}{2\lambda} \int_0^{\infty} \left[ f(-\sqrt{y}/\lambda) + f(\sqrt{y}/\lambda) \right] \frac{dy}{\sqrt{y}}$$

–4–

obtain the following J-matrix elements associated with the wave equation (3.1) expressed in a symmetric form

$$J_{m,n} = \langle \phi_m | (H-\varepsilon) | \phi_n \rangle = \frac{1}{2}\Big\{[4n+4\alpha+1-\varepsilon]\langle \phi_m | \phi_n \rangle - \langle \phi_m | y | \phi_n \rangle$$
$$+\Big[4n(v+\tfrac{1}{2}-2\alpha)-2\alpha(2\alpha-1)\Big]\langle \phi_m | \tfrac{1}{y} | \phi_n \rangle + \langle \phi_m | U(y) | \phi_n \rangle \quad (3.2)$$
$$+4\Big(2\alpha-v-\tfrac{1}{2}\Big)\tfrac{A_n}{A_{n-1}}(n+v)\langle \phi_m | \tfrac{1}{y} | \phi_{n-1} \rangle \Big\} + n \leftrightarrow m$$

Requiring this to be a tridiagonal matrix dictates that the last term either be eliminated by choosing $2\alpha = v + 1/2$ or be proportional to $\delta_{m,n-1}$ which requires $2\alpha = v + 3/2$. Thus, we have two possibilities and we will study them separately.

**3.1 Case $2\alpha = v + ½$:**

Due to our original requirement that $\alpha \geq 0$ the above equality requires $v \geq -\frac{1}{2}$. The J-matrix elements, in this case, reduce to

$$J_{m,n} = 2\Big(2n+v+1-\tfrac{\varepsilon}{2}\Big)\langle \phi_m | \phi_n \rangle$$
$$+\Big(\tfrac{1}{4}-v^2\Big)\langle \phi_m | \tfrac{1}{y} | \phi_n \rangle - \langle \phi_m | y | \phi_n \rangle + \langle \phi_m | U(y) | \phi_n \rangle \quad (3.3)$$

Evaluation of the tridiagonal matrix elements gives

$$\left.\begin{array}{c}\langle \phi_m | \phi_n \rangle = \delta_{m,n} \\ \langle \phi_m | y | \phi_n \rangle = (2n+v+1)\delta_{m,n} - \sqrt{n(n+v)}\,\delta_{m,n-1} - \sqrt{(n+1)(n+v+1)}\,\delta_{m,n+1}\end{array}\right\} \quad (3.4)$$

whereas the second term in equation (3.3), $\langle \phi_m | \tfrac{1}{y} | \phi_n \rangle$, is not tridiagonal. This term could be eliminated either by choosing $v = \pm \tfrac{1}{2}$ or be cancelled by a counter term in the potential function $U(y)$. For the first choice, the solvable potential that maintains the tridiagonal structure is (up to a constant) equal to $U(y) = ay$. In terms of $x$, it is the Harmonic oscillator potential $U(x) = a\lambda^2 x^2$. Consequently, we obtain the following J-matrix elements

$$J_{m,n} = 2\Big(2n \pm \tfrac{1}{2}+1-\tfrac{\varepsilon}{2}\Big)\delta_{m,n}$$
$$+(a-1)\Big[\Big(2n \pm \tfrac{1}{2}+1\Big)\delta_{m,n} - \sqrt{n\Big(n \pm \tfrac{1}{2}\Big)}\,\delta_{m,n-1} - \sqrt{(n+1)\Big(n \pm \tfrac{1}{2}+1\Big)}\,\delta_{m,n+1}\Big] \quad (3.5)$$

The requirement that the J-matrix (3.5) be diagonal leads to the discrete eigen-energies and corresponding wavefunctions for the bound state. This requirement dictates that $a = 1$ and results in the following energy spectrum

$$E_n = \hbar\omega\begin{cases} 2n+\tfrac{1}{2} & ,v = -\tfrac{1}{2} \\ 2n+\tfrac{3}{2} & ,v = +\tfrac{1}{2} \end{cases} \quad (3.6)$$

where the oscillator frequency $\omega = \hbar\lambda^2/m$. The corresponding normalized wave functions are connected to the even and odd Hermite polynomials:

$$\Psi_n(x) = \phi_n(x) = \frac{(-1)^n \sqrt{\lambda}\, e^{-\lambda^2 x^2/2}}{2^{2n}\sqrt{\Gamma(n+1)\Gamma(n+3/2)}}\begin{cases} \sqrt{n+\tfrac{1}{2}}\,H_{2n}(\lambda x) & ;v=-\tfrac{1}{2} \\ H_{2n+1}(\lambda x) & ;v=+\tfrac{1}{2} \end{cases} \quad (3.7)$$

where we have used the well-known relation of the even and odd Hermite polynomials to $L_n^{-\frac{1}{2}}(x^2)$ and $xL_n^{\frac{1}{2}}(x^2)$, respectively. Now, when $v \neq \pm 1/2$ we can still eliminate the non tridiagonal term in equation (3.3) by requiring the presence of a counter term in the potential. Thus the solvable potential in this case is given by:

$$U(y) = ay + \frac{b}{y} \quad \rightarrow \quad U(x) = Ax^2 + \frac{B}{x^2}; \quad A = a\lambda^2 \quad ; \quad B = b/\lambda^2 \quad (3.8)$$

where $b = v^2 - \frac{1}{4}$, $b > -\frac{1}{4}$ and $b \neq 0$. The potential given in (3.8) is the harmonic oscillator added to the inverse square potential in the original $x$ variable. Therefore, the basis parameter $v$ depend on the potential parameter $b$. Using the transformation

$$f_n = \sqrt{\frac{\Gamma(n+1)}{\lambda \Gamma(n+v+1)}} \, d_n, \quad (3.9)$$

we obtain the following recursion relation

$$[(a+1)(2n+v+1) - \varepsilon] d_n - (a-1)[(n+v) d_{n-1} + (n+1) d_{n+1}] = 0 \, . \quad (3.10)$$

This equation can be written in different forms depending on the range of the potential parameter $a$. When $a > 1$ we can write (3.10) as follows:

$$2\left[\left(n + \frac{v+1}{2} - \frac{\varepsilon}{4}\right)\cosh\theta + \frac{\varepsilon}{4}\right] d_n = (n+v) d_{n-1} + (n+1) d_{n+1}, \quad (3.11)$$

where $\cosh\theta = \frac{a+1}{a-1}$ and $\theta > 0$. The solution of the above recursion relation is given by the Hyperbolic Pollaczek polynomials defined by (A.14) in the Appendix. Thus, the un-normalized stationary wave function is then given by

$$\Psi(x, \varepsilon) = |\lambda x|^{v+\frac{1}{2}} e^{-\lambda^2 x^2/2} \sum_{n=0}^{\infty} \frac{\Gamma(n+1)}{\Gamma(n+v+1)} \, \tilde{P}_n^{\frac{v+1}{2}}\left(\frac{a+1}{a-1}; -\frac{\varepsilon}{4}, \frac{\varepsilon}{4}\right) L_n^v(\lambda^2 x^2), \quad (3.12)$$

where $A \geq 0$. Similarly, the wave function for $0 < a < 1$ and $a < 0$ are given by

$$\Psi(x, \varepsilon) = |\lambda x|^{v+\frac{1}{2}} e^{-\lambda^2 x^2/2} \sum_{n=0}^{\infty} \frac{\Gamma(n+1)}{\Gamma(n+v+1)} L_n^v(\lambda^2 x^2) \begin{cases} (-)^n \tilde{P}_n^{\frac{v+1}{2}}\left(\frac{1+a}{1-a}; -\frac{\varepsilon}{4}, -\frac{\varepsilon}{4}\right) & ; 0 < a < 1 \\ P_n^{\frac{v+1}{2}}\left(\frac{a+1}{a-1}; -\frac{\varepsilon}{4}, \frac{\varepsilon}{4}\right) & ; a < 0 \end{cases} \quad (3.13)$$

Requiring that equation (3.12) be diagonal dictates that $a = 1$ and gives the eigen-energies

$$E_n = \hbar\omega\left[2n + 1 \pm \sqrt{\frac{1}{4} + b}\right], \quad (3.14)$$

where the minus sign holds only for $b$ in the range $-\frac{1}{4} < b < 0$.

**3.2 Case 2 $\alpha = v + 3/2$ :**

In this case the J-matrix becomes

$$J_{m,n} = 2\left(n + m + v + 2 - \frac{\varepsilon}{2}\right)\langle\phi_m|\phi_n\rangle - \langle\phi_m|y|\phi_n\rangle - \left[4n + \left(v + \frac{3}{2}\right)\left(v + \frac{1}{2}\right)\right]\langle\phi_m|\frac{1}{y}|\phi_n\rangle \\ + 2\sqrt{n(n+v)} \, \langle\phi_m|\frac{1}{y}|\phi_{n-1}\rangle + 2\sqrt{m(m+v)} \, \langle\phi_n|\frac{1}{y}|\phi_{m-1}\rangle + \langle\phi_m|U(y)|\phi_n\rangle \quad (3.15)$$

$\langle\phi_m|y|\phi_n\rangle$ is the only non-tridiagonal term and hence should be eliminated by a counter term in $U(y)$ that could include additionally a $1/y$ term without violating tridiagonalization. Hence the general form of the solvable potential in this case reads.

$$U(y) = y + \frac{b}{y} \quad \rightarrow \quad U(x) = Ax^2 + \frac{B}{x^2}; \quad A = \lambda^2 \quad ; \quad B = b/\lambda^2 \quad (3.16)$$

where $b$ is an arbitrary constant. The resulting three-term recursion relation for the expansion coefficients of the wavefunction is

$$\left[(n+v+1)\left(n+\tfrac{v}{2}+1-\tfrac{\varepsilon}{4}\right)+n\left(n+\tfrac{v}{2}-\tfrac{\varepsilon}{4}\right)-\left(\tfrac{v+1}{2}\right)^2+\tfrac{b}{4}+\tfrac{1}{16}\right]d_n$$
$$-n\left(n+\tfrac{v}{2}-\tfrac{\varepsilon}{4}\right)d_{n-1}-(n+v+1)\left(n+\tfrac{v}{2}+1-\tfrac{\varepsilon}{4}\right)d_{n+1}=0 \qquad (3.17)$$

where $d_n$ is related to the expansion coefficients $f_n$ by the relation (3.9). We compare this recursion relation to that of the continuous dual Hahn orthogonal polynomials (A.17) in the Appendix. Consequently, we obtain

$$f_n(\varepsilon)=\sqrt{\frac{\Gamma(n+1)}{\lambda\Gamma(n+v+1)}}\,S_n^{\frac{v+1}{2}}\!\left(-\tfrac{4b+1}{16};\tfrac{v+1}{2},\tfrac{2-\varepsilon}{4}\right). \qquad (3.18)$$

Moreover, reality of the representation demands that $b\leq-\tfrac{1}{4}$. In fact, it is well known that the dimensionless coupling parameter of the inverse square potential must satisfy this same condition if bound state solutions were to exist [9]. The un-normalized wave function becomes

$$\Psi(x,\varepsilon)=(\lambda x)^{v+\frac{3}{2}}e^{-\lambda^2 x^2/2}\sum_{n=0}^{\infty}\frac{\Gamma(n+1)}{\Gamma(n+v+1)}\,S_n^{\frac{v+1}{2}}\!\left(-\tfrac{4b+1}{16};\tfrac{v+1}{2},\tfrac{2-\varepsilon}{4}\right)L_n^{v}(\lambda^2 x^2) \qquad (3.19)$$

## 4. The Morse Transformation

It is defined through the transformation $y=\mu e^{-\lambda x}$ with $x\in\mathbb{R}$ where $\mu$ and $\lambda$ are positive parameters with $\mu$ dimensionless and $\lambda$ having the dimension of inverse length. In this case the variable space is $[0,\infty[$ and the original Schrödinger equation is transformed to

$$\left[-y^2\frac{d^2}{dy^2}-y\frac{d}{dy}+U(y)-\varepsilon\right]\Psi(y)=0 \quad;\quad y\in[0,\infty[ \qquad (4.1)$$

The integration measure is given by: $\int_{-\infty}^{+\infty}dx\cdots=\int_{0}^{+\infty}\frac{dy}{\lambda y}\cdots$. Using the appropriate basis for this problem, which is the Laguerre basis in Eq. (2.2), we obtain the following $J$-matrix elements

$$J_{m,n}=\langle\phi_m|(H-\varepsilon)|\phi_n\rangle=\left[n(v-2\alpha)-\alpha^2-\varepsilon\right]\langle\phi_m|\phi_n\rangle+\left(n+\alpha+\tfrac{1}{2}\right)\langle\phi_m|y|\phi_n\rangle$$
$$-\tfrac{1}{4}\langle\phi_m|y^2|\phi_n\rangle+\langle\phi_m|U(y)|\phi_n\rangle+(n+v)(2\alpha-v)\frac{A_n}{A_{n-1}}\langle\phi_m|\phi_{n-1}\rangle \qquad (4.2)$$

which is required to be tridiagonal by construction. Following the same approach as in section 3, we get two possibilities; $v=2\alpha$ or $v=2\alpha-1$. As an example, we consider only the first possibility. In this case the J-matrix element simplifies to

$$J_{m,n}=-\left(\tfrac{v^2}{4}+\varepsilon\right)\langle\phi_m|\phi_n\rangle+\langle\phi_m|U(y)|\phi_n\rangle+\left(n+\tfrac{v+1}{2}\right)\langle\phi_m|y|\phi_n\rangle-\tfrac{1}{4}\langle\phi_m|y^2|\phi_n\rangle \quad (4.3)$$

Evaluation of the above matrix elements gives

$$\left.\begin{array}{c}\langle\phi_m|y|\phi_n\rangle=\delta_{m,n}\\ \langle\phi_m|y^2|\phi_n\rangle=(2n+v+1)\delta_{m,n}-\sqrt{n(n+v)}\,\delta_{m,n-1}-\sqrt{(n+1)(n+v+1)}\,\delta_{m,n+1}\end{array}\right\} \qquad (4.4)$$

However, the first term in (4.3) is not tridiagonal and should be eliminated. This term could not be cancelled by a counter term in the potential if we do not prefer to deal with the issue of energy dependent potentials in this setting. Therefore, we choose the alternative, which is requiring its factor to vanish leading to $\varepsilon=-v^2/4$ (i.e., negative energy corresponding to bound states). This, however, makes the basis energy dependent. Now, the exactly solvable potential that conforms to the tridiagonal matrix representation of the wave operator can be written as follows

$$U(y) = a y + b y^2 \to V(x) = A e^{-\lambda x} + B e^{-2\lambda x}; \quad a = A/\mu, b = B/\mu^2, \quad (4.5)$$

where $A$ and $B$ are arbitrary constants. This is a generalized version of the Morse potential [10]. The corresponding three-term recursion relation for the expansion coefficient of the wave function is given by

$$2\left[\left(b+\tfrac{1}{4}\right)\left(n+\tfrac{\nu+1}{2}\right) + \tfrac{a}{2}\right] d_n = \left(b-\tfrac{1}{4}\right)\left[(n+\nu) d_{n-1} + (n+1) d_{n+1}\right] \quad (4.6)$$

where $d_n$ is defined in Eq. (3.9). For $b < 0$ (i.e., for negative potential parameter $B$) this could be written as

$$2\left[\left(n+\tfrac{\nu+1}{2}+a\right)\cos\theta - a\right] d_n = (n+\nu) d_{n-1} + (n+1) d_{n+1}, \quad (4.7)$$

where $\cos\theta = \tfrac{b+1/4}{b-1/4}$. Comparing this with the recursion relation of the Pollaczek polynomial (A.11), we obtain the following solution

$$\Psi(y,\varepsilon) = y^{\nu/2} e^{-y/2} \sum_{n=0}^{\infty} \tfrac{\Gamma(n+1)}{\Gamma(n+\nu+1)} P_n^{\frac{\nu+1}{2}}\left(\tfrac{b+1/4}{b-1/4}; a, -a\right) L_n^\nu(y), \quad (4.8)$$

where $\nu$ is energy dependent and given as $\nu = 2\sqrt{-\varepsilon}$. If the potential parameter $B$ is positive then we obtain the following solutions

$$\Psi(y,\varepsilon) = y^{\nu/2} e^{-y/2} \sum_{n=0}^{\infty} \tfrac{\Gamma(n+1)}{\Gamma(n+\nu+1)} L_n^\nu(y) \times \begin{cases} \tilde{P}_n^{\frac{\nu+1}{2}}\left(\tfrac{b+1/4}{b-1/4}; a, -a\right) &, b > \tfrac{1}{4} \\ (-)^n \tilde{P}_n^{\frac{\nu+1}{2}}\left(\tfrac{1+b}{\tfrac{1}{4}-b}; a, a\right) &, b < \tfrac{1}{4} \end{cases} \quad (4.9)$$

The bound states are given by the diagonalization requirements obtained from (4.6) as

$$b - \tfrac{1}{4} = 0, \quad n + \tfrac{\nu+1}{2} + a = 0, \quad (4.10)$$

with $\varepsilon = -\nu^2/4$ (from above) and $\nu > -1$. Since $b = 1/4$ is positive, the potential given in (4.5) is a parabola with a minimum $U_{\min} = -a^2$ at $y = -a/2b = -2a$ this requires that $a$ be negative (i.e., $A < 0$). Therefore, we obtain the following finite bound states energy spectrum

$$\varepsilon_n = -\tfrac{\nu^2}{4} = -\left(n + a + \tfrac{1}{2}\right)^2 \quad n = 0, 1, \ldots, n_{\max} \quad (4.11).$$

The fact that the energy cannot be less than $U_{\min}$ imposes a limit on the number of possible bound states, $n_{\max}$, where $n_{\max}$ is the maximum integer less than or equal to $-2a - \tfrac{1}{2}$. The corresponding bound state wave functions are given by:

$$\phi_n(y) = A_n y^{\nu/2} e^{-y/2} L_n^\nu(y), \quad (4.12)$$

where $\nu = \pm 2(n + a + 1/2)$ and the choice of sign should comply with $\nu > -1$.

## 5. Rosen-Morse Transformation

It is defined through the transformation $y = \tanh(\lambda x)$ where $x \in \mathbb{R}$ and $\lambda$ is a positive length scale parameter. In this case the variable space is $[-1, +1]$ and the Schrödinger equation becomes

$$\left[ -(1-y^2)\frac{d}{dy}(1-y^2)\frac{d}{dy} + U(y) - \varepsilon \right] \Psi(y,\varepsilon) = 0 \quad ; \quad y \in [-1,1] \tag{5.1}$$

The integration measure is defined by

$$\int_{-\infty}^{+\infty} dx \cdots = \int_{-1}^{+1} \frac{dy}{\lambda(1-y^2)} \cdots \tag{5.2}$$

Using the Jacobi basis given by Eq. (2.3), we obtain the *J*-matrix element. It includes, besides the potential matrix elements, the following terms $\langle \phi_m | \phi_n \rangle$, $\langle \phi_m | (1 \pm y) | \phi_n \rangle$, $\langle \phi_m | (1 - y^2) | \phi_n \rangle$, and $\langle \phi_m | (1 \pm y) | \phi_{n-1} \rangle$. Imposing the tridiagonal requirement, one is lead to three possibilities for the basis parameters: $(\alpha,\beta) = \left(\frac{\nu}{2},\frac{\mu}{2}\right)$, $\left(\frac{\nu+1}{2},\frac{\mu}{2}\right)$ or $\left(\frac{\nu}{2},\frac{\mu+1}{2}\right)$. The second and third cases are easily obtained from each other by exchanging the parameter $\mu$ and $\nu$, we also consider only the second case where $(\alpha,\beta) = \left(\frac{\nu+1}{2},\frac{\mu}{2}\right)$. After some manipulations we can express the J-matrix as follows

$$\begin{aligned} J_{m,n} &= \langle \phi_m | (1-y) | \phi_n \rangle \left[ -2\frac{n(n+\mu)}{2n+\mu+\nu} - \frac{1}{2}(\mu+\nu+1)(\nu-\mu+1) \right] + \langle \phi_m | U(y) | \phi_n \rangle \\ &+ \frac{1}{8}\left[ (2n+\mu+\nu+2)^2 \langle \phi_m | (1-y^2) | \phi_n \rangle + n \leftrightarrow m \right] - (\mu^2 + \varepsilon) \langle \phi_m | \phi_n \rangle \\ &- \left[ \frac{1}{2n+\mu+\nu} \sqrt{\frac{n(n+\mu)(n+\nu)(n+\mu+\nu)(2n+\mu+\nu+1)}{2n+\mu+\nu-1}} \langle \phi_m | (1-y) | \phi_{n-1} \rangle + n \leftrightarrow m \right] \end{aligned} \tag{5.3}$$

Eliminating the non tridiagonal term $\langle \phi_m | \phi_n \rangle$ require that $\varepsilon = -\mu^2$ which means that the basis parameter $\mu$ depends on the energy. The solvable model potential will be allowed to contain all contributions leading to a tridiagonal representation of the *J*-matrix element. Hence we can write

$$U(y) = A(1-y) + B(1-y^2) \tag{5.4}$$

which can be written in terms of the original *x* variable as follows

$$V(x) = A - A\tanh(\lambda x) + \frac{B}{\cosh^2(\lambda x)}, \tag{5.5}$$

where *A* and *B* are arbitrary real potential parameters. The recursion relation for the expansion coefficients of the wave function associated with the above *J*-matrix reads as follows

$$\begin{aligned} A\, d_n &= \Big\{ \tfrac{1}{2}(\nu+\mu+1)(\nu-\mu+1) + \tfrac{2n(n+\mu)}{2n+\mu+\nu} \\ &\quad + \left[ \tfrac{\mu^2-\nu^2}{(2n+\mu+\nu)(2n+\mu+\nu+2)} - 1 \right] \left[ B - \tfrac{1}{4} + \tfrac{1}{4}(2n+\mu+\nu+2)^2 \right] \Big\} d_n \\ &\quad + \tfrac{2(n+1)(n+\mu+\nu+1)}{(2n+\mu+\nu+1)(2n+\mu+\nu+2)} \left[ B - \tfrac{1}{4} + \tfrac{1}{4}(2n+\mu+\nu+2)^2 \right] d_{n+1} \\ &\quad + \tfrac{2(n+\mu)(n+\nu)}{(2n+\mu+\nu)(2n+\mu+\nu+1)} \left[ B - \tfrac{1}{4} + \tfrac{1}{4}(2n+\mu+\nu)^2 \right] d_{n-1} \end{aligned} \tag{5.6}$$

The polynomial solutions of this recursion do not match any of the classical orthogonal polynomials. However, they are completely specified to all orders by choosing the standard initial value $d_0 = 1$. The discrete spectrum is given by the diagonal requirement (1.3) which in this case gives

$$-B + \tfrac{1}{4} = \tfrac{1}{4}(2n+\mu+\nu+2)^2 \tag{5.7a}$$

$$A = \tfrac{1}{2}(\nu+\mu+1)(\nu-\mu+1) + \tfrac{2n(n+\mu)}{2n+\mu+\nu} \tag{5.7b}$$

The first equation requires that $B \le \frac{1}{4}$. These two equations could be solved for the two parameters $\mu$ and $\nu$ in terms of $A$, $B$, and $n$. Therefore, the energy spectrum will be determined in terms of the potential parameters and the index $n$ as

$$\varepsilon_n = -\frac{1}{4}\left[\gamma - 2(n+1) + \frac{\gamma-2}{\gamma-2(n+1)}\left(1 - \frac{2A}{\gamma-1}\right)\right]^2 \tag{5.8}$$

where $\gamma = \sqrt{-B + 1/4}$.

## 6. Conclusion

With the help of a regular mapping $y(x)$ which defines a $y$-physical domain as the image of the original $x$-physical domain we have been able to obtain a closed form solution of the Schrödinger equation for a given potential $V(x)$ by requiring a tridiagonal representation of the Schrödinger operator. With this approach we generated a large class of regular solutions of the original Schrödinger equation. The problem then reduces to finding solutions of the resulting three term recursion relation for the expansion coefficient of the wave function in a suitable square integrable basis set. Usually the solutions of the three term recursion relations are either expressed in terms a variant of the classical orthogonal polynomials or one can express physical quantities in terms of Green function which in their turn are expressed in terms of continued fractions [11]. The bound states and their related wave functions constitute a sub-class generated by requiring a diagonal representation.

Finally we note that the examples presented in this work do not exhaust all possible potentials in the class of 1D analytically solvable potentials. The limitations of our approach are reflected in the limited type of mappings $y(x)$ that enable the y-physical domain to be restricted to either finite or semi-infinite intervals so as to enable us to use suitable orthogonal polynomials as a basis set. Nonetheless, all known classes of exactly solvable potentials fall within the range of application of our approach. In fact to avoid expanding this work out of proportion we did not include few other mappings that generate other solvable potentials such as Poschl-Teller, modified Poschl-Teller and general power law potentials which are amenable to our approach. This work can be considered as an extension of the previous work done by one of the authors [12] and made more readable for a general audience. Moreover, it has also been applied recently to study quasi-exactly solvable potentials in the non-relativistic case [13] and will also be extended to the relativistic domain in the near future.

**Acknowledgments**

We acknowledge the support of the Physics Department at King Fahd University of Petroleum and Minerals under project FT-2006-05.

**Appendix A.**

The following are useful formulas and relations satisfied by the orthogonal polynomials that are relevant to the development carried out in this work. They are found in most books on orthogonal polynomials [14]. We list them here for ease of reference.

(1) The Laguerre polynomials $L_n^\nu(x)$, where $\nu > -1$:

$$xL_n^\nu = (2n+\nu+1)L_n^\nu - (n+\nu)L_{n-1}^\nu - (n+1)L_{n+1}^\nu \tag{A.1}$$

$$L_n^\nu(x) = \frac{\Gamma(n+\nu+1)}{\Gamma(n+1)\Gamma(\nu+1)} {}_1F_1(-n;\nu+1;x) \tag{A.2}$$

$$\left[x\frac{d^2}{dx^2} + (\nu+1-x)\frac{d}{dx} + n\right]L_n^\nu(x) = 0 \tag{A.3}$$

$$x\frac{d}{dx}L_n^\nu = nL_n^\nu - (n+\nu)L_{n-1}^\nu \tag{A.4}$$

$$\int_0^\infty x^\nu e^{-x} L_n^\nu(x) L_m^\nu(x) dx = \frac{\Gamma(n+\nu+1)}{\Gamma(n+1)} \delta_{nm} \tag{A.5}$$

(2) The Jacobi polynomials $P_n^{(\mu,\nu)}(x)$, where $\mu > -1, \nu > -1$:

$$\left(\frac{1\pm x}{2}\right) P_n^{(\mu,\nu)} = \frac{2n(n+\mu+\nu+1)+(\mu+\nu)(\frac{\mu+\nu}{2} \pm \frac{\nu-\mu}{2}+1)}{(2n+\mu+\nu)(2n+\mu+\nu+2)} P_n^{(\mu,\nu)}$$
$$\pm \frac{(n+\mu)(n+\nu)}{(2n+\mu+\nu)(2n+\mu+\nu+1)} P_{n-1}^{(\mu,\nu)} \pm \frac{(n+1)(n+\mu+\nu+1)}{(2n+\mu+\nu+1)(2n+\mu+\nu+2)} P_{n+1}^{(\mu,\nu)} \tag{A.6}$$

$$P_n^{(\mu,\nu)}(x) = \frac{\Gamma(n+\mu+1)}{\Gamma(n+1)\Gamma(\mu+1)} {}_2F_1(-n, n+\mu+\nu+1; \mu+1; \frac{1-x}{2}) = (-)^n P_n^{(\nu,\mu)}(-x) \tag{A.7}$$

$$\left\{(1-x^2)\frac{d^2}{dx^2} - [(\mu+\nu+2)x + \mu-\nu]\frac{d}{dx} + n(n+\mu+\nu+1)\right\} P_n^{(\mu,\nu)}(x) = 0 \tag{A.8}$$

$$(1-x^2)\frac{d}{dx} P_n^{(\mu,\nu)} = -n\left(x + \frac{\nu-\mu}{2n+\mu+\nu}\right) P_n^{(\mu,\nu)} + 2\frac{(n+\mu)(n+\nu)}{2n+\mu+\nu} P_{n-1}^{(\mu,\nu)} \tag{A.9}$$

$$\int_{-1}^{+1} (1-x)^\mu (1+x)^\nu P_n^{(\mu,\nu)}(x) P_m^{(\mu,\nu)}(x) dx = \frac{2^{\mu+\nu+1}}{2n+\mu+\nu+1} \frac{\Gamma(n+\mu+1)\Gamma(n+\nu+1)}{\Gamma(n+1)\Gamma(n+\mu+\nu+1)} \delta_{nm} \tag{A.10}$$

(3) The Pollaczek polynomials $P_n^\mu(x;a,b)$, where $\mu > 0$, $a \geq |b|$, $x = \cos\theta$, and $0 < \theta < \pi$:

$$2[(n+\mu+a)x + b] P_n^\mu = (n-1+2\mu) P_{n-1}^\mu + (n+1) P_{n+1}^\mu \tag{A.11}$$

$$P_n^\mu(x;a,b) = \frac{\Gamma(n+2\mu)}{\Gamma(n+1)\Gamma(2\mu)} e^{in\theta} {}_2F_1(-n, \mu+iy; 2\mu; 1-e^{-2i\theta}) \tag{A.12}$$

$$\int_{-1}^{+1} \rho^\mu(x) P_n^\mu(x;a,b) P_m^\mu(x;a,b) dx = \frac{\Gamma(n+2\mu)}{(n+\mu+a)\Gamma(n+1)} \delta_{nm}, \tag{A.13}$$

where $y = \frac{b+a\cos\theta}{\sin\theta}$ and $\rho^\mu(x) = \frac{1}{\pi}(2\sin\theta)^{2\mu-1} e^{(2\theta-\pi)y} |\Gamma(\mu+iy)|^2$.

(4) The Hyperbolic Pollaczek polynomials $\tilde{P}_n^\mu(x;a,b)$ is obtained from the Pollaczek polynomials by the transformation $\theta \to i\theta$ (i.e., $x = \cosh\theta$) and $\theta > 0$:

$$2[(n+\mu+a)x + b] \tilde{P}_n^\mu = (n-1+2\mu) \tilde{P}_{n-1}^\mu + (n+1) \tilde{P}_{n+1}^\mu \tag{A.14}$$

$$\tilde{P}_n^\mu(x;a,b) = \frac{\Gamma(n+2\mu)}{\Gamma(n+1)\Gamma(2\mu)} e^{-n\theta} {}_2F_1(-n,\mu+z;2\mu;1-e^{2\theta}) \tag{A.15}$$

$$\int_1^\infty \tilde{\rho}^\mu(x) \tilde{P}_n^\mu(x;a,b) \tilde{P}_m^\mu(x;a,b) dx = \frac{\Gamma(n+2\mu)}{(n+\mu+a)\Gamma(n+1)} \delta_{nm}, \tag{A.16}$$

where $z = \frac{b+a\cosh\theta}{\sinh\theta}$ and $\tilde{\rho}^\mu(x) = \frac{1}{\pi}(2i\sinh\theta)^{2\mu-1} e^{(2\theta-i\pi)z} |\Gamma(\mu+z)|^2$.

(5) The continuous dual Hahn polynomials $S_n^\mu(x^2;a,b)$, where $x$ is real and $\mu$, $a$, $b$ are positive except for a pair of complex conjugates with positive real parts:

$$\begin{aligned} x^2 S_n^\mu &= \left[ (n+\mu+a)(n+\mu+b) + n(n+a+b-1) - \mu^2 \right] S_n^\mu \\ &\quad - n(n+a+b-1) S_{n-1}^\mu - (n+\mu+a)(n+\mu+b) S_{n+1}^\mu \end{aligned} \tag{A.17}$$

$$S_n^\mu(x^2;a,b) = {}_3F_2\left( \begin{array}{c} -n, \mu+ix, \mu-ix \\ \mu+a, \mu+b \end{array} \Big| 1 \right) \tag{A.18}$$

$$\int_0^\infty \rho^\mu(x) S_n^\mu(x^2;a,b) S_m^\mu(x^2;a,b) dx = \frac{\Gamma(n+1)\Gamma(n+a+b)}{\Gamma(n+\mu+a)\Gamma(n+\mu+b)} \delta_{nm} \tag{A.19}$$

where $\rho^\mu(x) = \frac{1}{2\pi} \left| \frac{\Gamma(\mu+ix)\Gamma(a+ix)\Gamma(b+ix)}{\Gamma(\mu+a)\Gamma(\mu+b)\Gamma(2ix)} \right|^2$.

**References**

[1] See, for example, R. De, R. Dutt and U. Sukhatme, J. Phys. A25, L843 (1992); B. Bagchi and A. Ganguly, J. Phys. A36, L161 (2003); L.D. Salem and R. Montemayor, Phys. Rev. A43, 1169 (1991); J. W. Dabrowska, Avinash Khare ands Uday P. Sukhatme, J. Phys. A 21, L 195 (1988); G. A. Natanzon, Teor. Mat. Fiz. **38**, 219 (1979) [Theor. Math. Phys. **38**, 146 (1979)]; L. E. Gendenshtein, Zh. Eksp. Teor. Fiz. Pis'ma Red. **38**, 299 (1983) [JETP Lett. **38**, 356 (1983)]; F. Cooper, J. N. Ginocchio, and A. Khare, Phys. Rev. D **36**, 2458 (1987); R. Dutt, A. Khare, and U. P. Sukhatme, Am. J. Phys. **56**, 163 (1988); **59**, 723 (1991); G. Lévai, J. Phys. A **22**, 689 (1989); **27**, 3809 (1994).

[2] S. Flugge, " Practical Quantum Mechanics", Springer-Verlag, New York, Heidelberg, Berlin (1974); L. D. Landau and E. M. Lifshitz, Quantum Mechanics, Pergamon Press Ltd. (1977).

[3] A. de Souza-Dutra, Phys. Rev. A **47**, R2435 (1993); N. Nag, R. Roychoudhury, and Y. P. Varshni, Phys. Rev. A **49**, 5098 (1994); R. Dutt, A. Khare, and Y. P. Varshni, J. Phys. A **28**, L107 (1995); C. Grosche, J. Phys. A, **28**, 5889 (1995); **29**, 365 (1996); G. Lévai and P. Roy, Phys. Lett. A **270**, 155 (1998); G. Junker and P. Roy, Ann. Phys. (N.Y.) **264**, 117 (1999)

[4] For more recent developments, see for example, B. Bagchi and C. Quesne, J. Phys. A **37** L133 (2004); A. Sinha, G. Lévai and P. Roy, Phys. Lett. A **322**, 78 (2004); B. Chakrabarti and T. K. Das, Mod. Phys. Lett. A **17**, 1367 (2002); R. Roychoudhury, P. Roy, M. Zonjil, and G. Lévai, J. Math. Phys. **42**, 1996 (2001)

[5] A. V. Turbiner, Commun. Math. Phys. **118**, 467 (1988); M. A. Shifman, Int. J. Mod. Phys. A **4**, 2897 (1989); R. Adhikari, R. Dutt, and Y. Varshni, Phys. Lett. A **141**, 1 (1989); J. Math. Phys. **32**, 447 (1991); R. Roychoudhury, Y. P. Varshni, and M. Sengupta, Phys. Rev. A **42**, 184 (1990); L. D. Salem and R. Montemayor, Phys. Rev. A **43**, 1169 (1991); M. W. Lucht and P. D. Jarvis, Phys. Rev. A **47**, 817

(1993); A. G. Ushveridze, *Quasi-exactly Solvable Models in Quantum Mechanics* (IOP, Bristol, 1994)

[6] For more recent developments, see for example, N. Debergh and B. Van den Bossche, Int. J. Mod. Phys. A **18**, 5421 (2003); R. Atre and P. K. Panigrahi, Phys. Lett. A **317**, 46 (2003); B. Bagchi and A. Ganguly, J. Phys. A **36**, L161 (2003); V. M. Tkachuk and T. V. Fityo, Phys. Lett. A **309**, 351 (2003); Y. Brihaye and B. Hartmann, Phys. Lett. A **306**, 291 (2003); A. Ganguly, J. Math. Phys. **43**, 5310 (2002); R. Koc, M. Koca, and E. Korcuk, J. Phys. A **35**, L527 (2002); N. Debergh, J. Ndimubandi, and B. Van den Bossche, Ann. Phys. **298**, 361 (2002); N. Debergh, B. Van den Bossche, and B. F. Samsonov, Int. J. Mod. Phys. A **17**, 1577 (2002)

[7] R. W. Haymaker and L. Schlessinger, *The Padé Approximation in Theoretical Physics*, edited by G. A. Baker and J. L. Gammel (Academic Press, New York, 1970); N. I. Akhiezer, *The Classical Moment Problem* (Oliver and Boyd, Einburgh, 1965); D. G. Pettifor and D. L. Weaire (editors), *The Recursion Method and its Applications* (Springer-Verlag, Berlin, 1985); H. S. Wall, *Analytic Theory of Continued Fractions* (Chelsea Publishing, New York, 1948).

[8] E. J. Heller and H. A. Yamani, Phys. Rev. A **9**, 1201 (1974); H. A. Yamani and L. Fishman, J. Math. Phys. **16**, 410 (1975); A. D. Alhaidari, E. J. Heller, H. A. Yamani, and M. S. Abdelmonem (eds.), *The J-matrix method: Recent Developments and Selected Applications* (Springer, Heidelberg, 2007).

[9] Andrew M. Essin and David Griffiths, Am. J. Phys. 74, 109 (2006) and references therein.

[10] N. Rosen and P. M. Morse, Phys. Rev. 42, 210 (1932).

[11] H. Bahlouli, A.D. Al-Haidari and M. S. Abdelmonem, Phys. Letters A 367, 162 (2007).

[12] A. D. Alhaidari, Ann. Phys. (NY) **317**, 152 (2005).

[13] A. D. Alhaidari, J. Phys. A **40**, 6305 (2007).

[14] Examples of textbooks and monographs on special functions and orthogonal polynomials are: W. Magnus, F. Oberhettinger, and R. P. Soni, *Formulas and Theorems for the Special Functions of Mathematical Physics* (Springer-Verlag, New York, 1966); T. S. Chihara, *An Introduction to Orthogonal Polynomials* (Gordon and Breach, New York, 1978); G. Szegö, *Orthogonal polynomials*, 4$^{th}$ ed. (Am. Math. Soc., Providence, RI, 1997); R. Askey and M. Ismail, *Recurrence relations, continued fractions and orthogonal polynomials*, Memoirs of the Am. Math. Soc., Vol. 49 Nr. 300 (Am. Math. Soc., Providence, RI, 1984)